\documentclass[3p,times]{elsarticle}

\usepackage{ecrc}

\volume{00}

\firstpage{1}

\journalname{Physics Letters B}

\runauth{}

\jid{procs}

\jnltitlelogo{Physics Letters B}

\usepackage{amssymb}

\usepackage{amsthm}

\usepackage[figuresright]{rotating}

\begin{document}

\begin{frontmatter}

\dochead{}

\title{Alpha induced reaction cross section measurements on $^{162}$Er for the astrophysical $\gamma$ process}

\author[label1]{G.\,G.\,Kiss}

\ead{ggkiss@atomki.mta.hu, Fax: (+36)(52) 416 181}

\author[label1]{T.\,Sz\"ucs \footnote{Present Address: Helmholtz-Zentrum Dresden-Rossendorf (HZDR), 01328 Dresden, Germany}}

\author[label2,label3]{T.\,Rauscher}

\author[label1]{Zs.\,T\"or\"ok}

\author[label1]{Zs.\,F\"ul\"op}

\author[label1]{Gy.\,Gy\"urky}

\author[label1]{Z.\,Hal\'asz}

\author[label1]{E.\,Somorjai}

\address[label1]{Institute for Nuclear Research (MTA ATOMKI), H-4001 Debrecen, Hungary}
\address[label2]{Centre for Astrophysics Research, School of Physics, Astronomy and Mathematics, University of Hertfordshire, Hatfield AL10 9AB, United Kingdom}
\address[label3]{Department of Physics, University of Basel, 4052 Basel, Switzerland}

\begin{abstract}
The cross sections of the $^{162}$Er($\alpha$,$\gamma$)$^{166}$Yb and $^{162}$Er($\alpha$,n)$^{165}$Yb reactions have been measured for the first time. The radiative alpha capture reaction cross section was measured from E$_{c.m.}$ = 16.09 down to E$_{c.m.}$ = 11.21 MeV, close to the astrophysically relevant region (which lies between 7.8 and 11.48 MeV at 3 GK stellar temperature). The $^{162}$Er($\alpha$,n)$^{165}$Yb reaction was studied above the reaction threshold between E$_{c.m.}$ = 12.19 and 16.09 MeV. The fact that the $^{162}$Er($\alpha$,$\gamma$)$^{166}$Yb cross sections were measured below the ($\alpha$,n) threshold at first time in this mass region opens the opportunity to study directly the $\alpha$-widths required for the determination of astrophysical reaction rates. The data clearly show
that compound nucleus formation in this reaction proceeds differently than previously predicted.
\end{abstract}
\end{frontmatter}

\section{Introduction}
\label{int}

Low-energy ($\alpha$,$\gamma$) and ($\alpha$,n) measurements are of great interest for an improved determination of certain astrophysical reaction rates in $\gamma$-process nucleosynthesis. Photodisintegration of nuclei above Fe in explosive stellar processes (such as core-collapse supernovae or type Ia supernovae) is called $\gamma$-process \cite{woohow}. While the bulk of naturally occurring heavy nuclei is produced in neutron-capture processes \cite{kapgall,wallerstein}, about 35 proton-rich nuclides between Se and Hg are bypassed by these. Hypothetically, the $\gamma$-process could be responsible for 32 among these so-called p-nuclei, with other nucleosynthesis processes contributing to the remaining ones \cite{p-review}. The main problem of the $\gamma$-process is the production of the isotopes $^{92,94}$Mo and $^{96,98}$Ru which cannot be synthesized in core-collapse events in an amount observed in the Solar System. There may be further problems at mass numbers $150\leq A \leq 165$, although they are less pronounced. While the $\gamma$-process initially proceeds with ($\gamma$,n) reactions, at neutron numbers $N\geq 82$ ($\gamma$,$\alpha$) reactions can compete at proton-rich isotopes and lead to a deflection or branching in the synthesis path. 

\begin{table*}
\center
\caption{\label{tab:data} Available experimental database above the A $\approx$ 100 mass region which can be used to constrain the alpha widths at low energies (taken from the KADoNiS database \cite{kadonis}), the astrophysicaly relevant energy region --- calculated at T$_9$=3 GK --- \cite{rau_gamow} is indicated, too.}
\begin{tabular}{cccccc}
\hline
\multicolumn{1}{c}{target nucleus} &
\multicolumn{1}{c}{Gamow window}&
\multicolumn{1}{c}{($\alpha,\gamma$) energy range} &
\multicolumn{1}{c}{($\alpha$,n) energy range} &
\multicolumn{1}{c}{($\alpha$,n) threshold} &
\multicolumn{1}{c}{reference} \\
\multicolumn{1}{c}{} &
\multicolumn{1}{c}{[MeV]} &
\multicolumn{1}{c}{[MeV]} &
\multicolumn{1}{c}{[MeV]} &
\multicolumn{1}{c}{[MeV]} &
\multicolumn{1}{c}{} \\
\hline
$^{127}$I  & 6.21 - 8.64  & 9.50 - 15.15  & 9.62 - 15.15  & 7.97  &\cite{kis_i}\\
$^{130}$Ba & 6.82 - 10.17 & 11.61 - 16.00 & 12.05 - 16.00 & 10.81 &\cite{hal_ba}\\
$^{139}$La & 6.91 - 9.17  & 11.96 - 31.59 & 9.82 - 38.49  & 9.34  &\cite{ver_la}\\
$^{151}$Eu & 7.44 - 10.40 & 12.25 - 17.04 & 11.31 - 17.04 & 10.41 &\cite{gyu_eu}\\
$^{169}$Tm & 7.77 - 10.65 & 11.21 - 17.08 & 11.21 - 17.08 & 10.43 &\cite{kiss_plb}\\
$^{168}$Yb & 7.98 - 11.63 & 12.53 - 14.73 & 12.53 - 14.73 & 12.07 &\cite{net_yb}\\
\hline
$^{162}$Er & 7.80 - 11.48 & 11.21 - 16.09 & 12.18 - 16.09 & 11.98 &present work\\
\hline
\end{tabular}
\end{table*}

Theoretical studies of the nuclear uncertainties in the $\gamma$-process make use of large reaction networks with mainly theoretical reaction rates (taken from the Hauser-Feshbach (H-F) model \cite{hf}). They have shown that the reaction flow for the production of heavy p-nuclei (140 $\leq$ A $\leq$ 200) is strongly sensitive to the ($\gamma,\alpha$) photodisintegration rates \cite{rau06, rap06}. Experimental information about the most important $\gamma$-induced reactions can be obtained from the study of the inverse capture reactions and using the detailed balance theorem. This approach is not only technically less challenging, but also provides more relevant astrophysical
information than the direct study of the $\gamma$-induced reactions \cite{moh07, rau11,rau14}.
Recent experiments, however, indicate that the H-F predictions may overestimate the $\alpha$-capture cross sections at low energies by factor of 3 to 20 and the difference between the predictions and the experimental results are increasing with decreasing energies \cite{gyu10, sau11, kiss_plb} (it is worth to emphasize that the astrophysically relevant energy region, the so-called Gamow window, lies few MeV below the experimentally reachable energy region \cite{rau_gamow}). This would strongly impact the astrophysical reaction rates and through this affects the results of the $\gamma$-process reaction network studies. In summary, experimental data at low energies are urgently needed to confirm the path of the $\gamma$-process at mass numbers $150\leq A \leq 165$. 

The H-F cross section calculations are sensitive to different nuclear properties such as $\alpha$-, neutron-, $\gamma$- and proton-widths \cite{sensipaper}. At energies covered by the previous $\alpha$-induced reaction studies available above the $A\approx 100$ mass region (listed in Table \ref{tab:data}), the cross section predictions are not only sensitive to the $\alpha$-widths, but additionally to the $\gamma$- and neutron widths. Therefore, the extrapolation of the experimental data toward the astrophysically relevant energy region could be questionable since the impact of the different sensitivities on the cross section predictions have to be disentangled. 
However, at even lower energies, the picture changes, the uncertainty in the astrophysical ($\gamma$,$\alpha$) reaction rates is completely dominated by the uncertainty in the prediction of the subCoulomb $\alpha$ width, which is calculated using global alpha+nucleus optical potentials \cite{mcf,avri,deme,moh97,moh}. On one hand the parameters of the $\alpha$-nucleus optical potential can be derived in elastic alpha scattering experiments at energies roughly 5-8 MeV above the Gamow window \cite{kiss_in} and as a second step the parameters have to be extrapolated down to the astrophysically relevant energy region. On the other hand the subCoulomb $\alpha$ width can be probed in low-energy ($\alpha$,$\gamma$) and ($\alpha$,n) cross section measurements \cite{sau11, rau_tm}. Despite several attempts, the bulk of the present experimental data cannot be described consistently by any global $\alpha$+nucleus optical potential, yet. 

\begin{table}
\center
\caption{\label{tab:decay}Decay parameters of the $^{162}$Er($\alpha$,$\gamma$)$^{166}$Yb and $^{162}$Er($\alpha$,n)$^{165}$Yb (which decays by electron-capture to $^{165}$Tm) reaction products taken from the literature \cite{nds_166, nds_165} and calculated (marked with *) using the I(Tm K X-ray)/I(82.3 keV) ratio, available from \cite{1973_De_22, 1963_Ja_06}.}
\begin{tabular}{ccccc}
\hline
\multicolumn{1}{c}{Residual} &
\multicolumn{1}{c}{Half-} &
\multicolumn{1}{c}{Energy} &
\multicolumn{1}{c}{Relative } \\
\multicolumn{1}{c}{nucleus} &
\multicolumn{1}{c}{life [h]} &
\multicolumn{1}{c}{[keV]} &
\multicolumn{1}{c}{intensity [\%]} \\
\hline
$^{166}$Yb  &  56.7  $\pm$ 0.1   & 82.3 & 16.0 $\pm$ 0.7$^*$\\
$^{166}$Tm  &   7.70 $\pm$ 0.03  & 80.6 & 11.5 $\pm$ 0.9 \\
$^{165}$Tm  & 30.06  $\pm$ 0.03  & 242.9 & 35.5 $\pm$ 0.7\\
&&297.4 & 12.71 $\pm$ 0.25 \\
\hline
\end{tabular}
\end{table}

The present measurement of the $^{162}$Er($\alpha$,$\gamma$)$^{166}$Yb and $^{162}$Er($\alpha$,n)$^{165}$Yb reactions provides another important milestone in the test of the predicted $\alpha$ strengths at low energies. Not only consistently measured ($\alpha,\gamma$) and ($\alpha$,n) cross sections on the p-nucleus $^{162}$Er become available but for the first time in this mass region ($\alpha$,$\gamma$) cross sections also below the ($\alpha$,n) threshold --- where in the ($\alpha,\gamma$) H-F predictions among all widths only the $\alpha$-widths contribute --- become available. This fact was found to be essential for an unambiguous study of the $\alpha$ width and its energy dependence.

\section{Experimental approach}
\label{exp}

\begin{figure}
\center
\resizebox{0.42\columnwidth}{!}{\rotatebox{0}{\includegraphics[clip=]{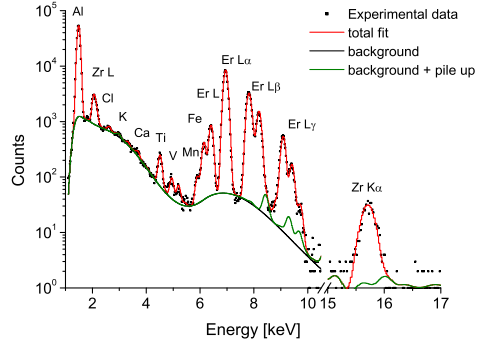}}}
\caption{\label{fig:pixe} PIXE spectrum measured by bombarding the Er targets with 2 MeV protons. The peaks used for the analysis
are marked. Peaks belonging to impurities in the target and/or the backing are indicated, too.}
\end{figure}

\begin{figure}
\center
\resizebox{0.74\columnwidth}{!}{\rotatebox{0}{\includegraphics[clip=]{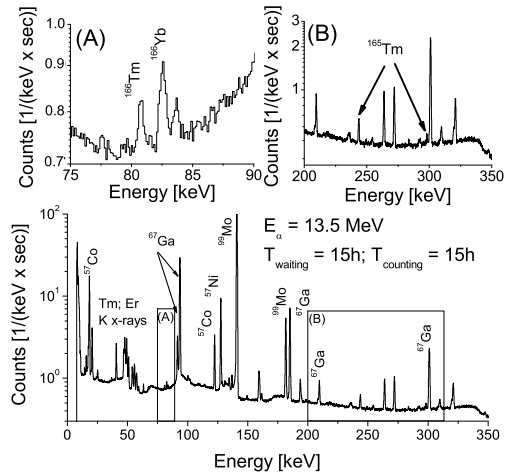}}}
\caption{\label{fig:spectra} Off-line $\gamma$ spectra normalized to the length of the countings (lower panel), taken after irradiating an Er target with 13.5 MeV $\alpha$ beam. The $\gamma$-lines used to determine the cross section of the $^{162}$Er($\alpha$,$\gamma$) (a) and $^{162}$Er($\alpha$,n) (b) reactions are marked.}
\end{figure}

The cross section measurement was carried out at the Institute for Nuclear Research of the Hungarian Academy of Sciences (MTA Atomki) using the activation technique. The electron capture decay of the Yb reaction products is followed by $\gamma$-ray emission which was detected using a Low Energy Photon Spectrometer (LEPS). The decay parameters of the investigated reactions are summarized in Table \ref{tab:decay}. In the next paragraphs a detailed description on the experiment can be found.

\begin{table*}
\center
\caption{\label{tab:res} Measured cross sections of the $^{162}$Er($\alpha$,$\gamma$)$^{166}$Yb and $^{162}$Er($\alpha$,n)$^{165}$Yb reactions.}
\begin{tabular}{ccccc}
\hline
\multicolumn{1}{c}{E$_{lab}$} &
\multicolumn{1}{c}{E$_{c.m.}$} &
\multicolumn{1}{c}{$^{162}$Er($\alpha$,$\gamma$)$^{166}$Yb} &
\multicolumn{1}{c}{$^{162}$Er($\alpha$,n)$^{165}$Yb} \\
\multicolumn{1}{c}{[MeV]} &
\multicolumn{1}{c}{[MeV]} &
\multicolumn{1}{c}{[$\mu$barn]} &
\multicolumn{1}{c}{[$\mu$barn]} \\
\hline
11.5 & 11.21 $\pm$ 0.06 & 1.10  $\pm$ 0.33 & \\
12.0 & 11.70 $\pm$ 0.07 & 2.50  $\pm$ 0.40 & \\
12.5 & 12.18 $\pm$ 0.07 & 6.57  $\pm$ 0.88 & 3.86 $\pm$ 0.82\\
13.0 & 12.67 $\pm$ 0.07 & 14.1  $\pm$ 1.6  & 12.2 $\pm$ 1.6\\
13.5 & 13.16 $\pm$ 0.06 & 30.2  $\pm$ 3.3  & 53.0 $\pm$ 5.6\\
14.0 & 13.65 $\pm$ 0.06 & 55.2  $\pm$ 5.9  & 161  $\pm$ 14\\
14.5 & 14.14 $\pm$ 0.07 & 106   $\pm$ 11   & 612  $\pm$ 66\\
15.5 & 15.10 $\pm$ 0.08 & 327   $\pm$ 35   & 3663 $\pm$ 345\\
16.5 & 16.09 $\pm$ 0.08 & 939   $\pm$ 102  & 17368 $\pm$ 1680\\
\hline
\end{tabular}
\end{table*}

The targets were made by reductive vacuum evaporation of Er$_2$O$_3$ powder enriched to 25.8\% in $^{162}$Er onto 2 $\mu$m thick, high purity Al foils. The Er$_2$O$_3$ powder was mixed with Zr powder and placed into a C crucible heated by electron beam. The absolute target thicknesses, the target impurities and the Zr contamination -- similar to \cite{kiss_plb} -- were determined using the PIXE technique \cite{pixe} and by X-ray fluorescence spectroscopy. The target thicknesses were found to be between 114 and 188 $\mu$g/cm$^2$ and the level of the Zr contamination was always below 4 atom\%. A typical PIXE spectrum can be seen in Fig. \ref{fig:pixe}.

The Er targets were then irradiated with $\alpha$ beams from the MGC cyclotron of MTA Atomki. The energy of the $\alpha$ beam was between E$_{lab}$ = 11.5 MeV and 16.5 MeV, this energy range was scanned with energy steps of 0.5 MeV - 1.0 MeV using beam currents of typically 2 $\mu$A. After the beam-defining aperture, the chamber was insulated and a secondary electron suppression voltage of $-300$ V was applied at the entrance of the chamber. The number of incident $\alpha$ particles in each irradiation was between 3.9 x 10$^{17}$ and 6.1 x 10$^{17}$. After the irradiations, T$_{waiting}$ = 0.25 h waiting time was used in order to let short-lived activities, which would impact the quality of the measurement, decay. The duration of the $\gamma$-countings were about 150-160 h in the case of each irradiation. To determine the $^{162}$Er($\alpha,\gamma$)$^{166}$Yb reaction cross section the yield of the 82.3 keV transition was measured. Furthermore, the $^{166}$Tm nucleus, the daughter of the produced unstable $^{166}$Yb decays by electron capture to $^{166}$Er with emission of 80.6 keV $\gamma$-ray, which was also used to determine the radiative $\alpha$ capture cross section. Since the half-life of $^{165}$Yb, produced by the $^{162}$Er($\alpha$,n) reaction, is relatively short, to determine the ($\alpha$,n) cross section on $^{162}$Er, the decay of its daughter ($^{165}$Tm nucleus) was investigated. A typical off-line $\gamma$ spectrum can be seen in Fig. \ref{fig:spectra}.

The uncertainty of the relative intensity of the 82.3 keV gamma transition is missing in \cite{nds_166}. In \cite{1973_De_22, 1963_Ja_06} the I(Tm K X-ray)/I(82.3 keV) ratio is given (8.68 $\pm$ 0.21 and 8.17 $\pm$ 0.23, respectively), the weighted average of these data (8.44 $\pm$ 0.27) together with the known X-ray intensities taken from \cite{nds_166} were used to calculate the value given in Table \ref{tab:decay}. The agreement between the cross sections based on the counting of the 82.3 keV and the 80.6 keV $\gamma$ rays where always within 3.4\%. 

\begin{figure}
\center
\resizebox{0.74\columnwidth}{!}{\rotatebox{0}{\includegraphics[clip=]{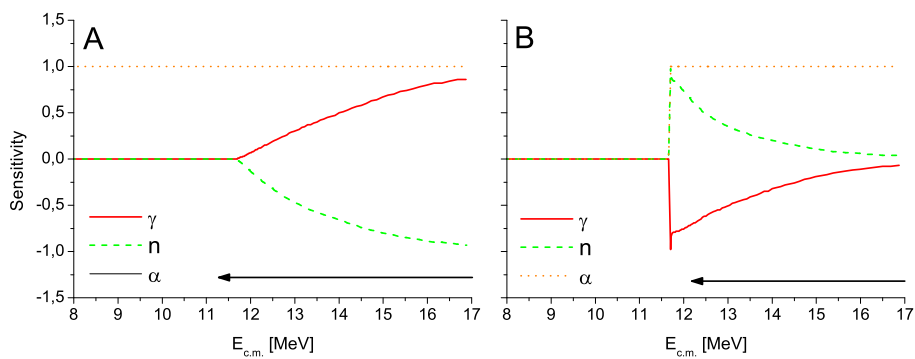}}}
\caption{\label{fig:sensi_ag} Sensitivities of the $^{162}$Er($\alpha$,$\gamma$)$^{166}$Yb (A) and $^{162}$Er($\alpha$,n)$^{165}$Yb (B) Hauser-Feshbach reaction cross sections to variations in the averaged neutron-, $\gamma$-, and $\alpha$-widths \cite{sensipaper}. The energy region where cross section data was measured are indicated by the arrows.}
\end{figure}

The low yields encountered in the present work necessitated the use of short source-to-detector distances for the $\gamma$-countings. The distance between the activated target and the Be window of the LEPS was 1 cm, the detector efficiencies had to be known in this geometry with high precision. For this purpose the following procedure was used: first the absolute detector efficiency was measured in far geometry: at 15 cm distance from the surface of the detector, using calibrated $^{57}$Co, $^{133}$Ba, $^{152}$Eu, and $^{241}$Am sources. Since the calibration sources (especially $^{133}$Ba, $^{152}$Eu) emit multiple $\gamma$-radiations from cascade transitions, in close geometry a strong true coincidence summing effect is expected resulting in an increased uncertainty of the measured efficiency. Therefore, no direct efficiency measurement in close geometry has been carried out. Instead, in the case of the high energy irradiations (at and above 14.0 MeV) the yield of the investigated $\gamma$-rays has been measured both in close and far geometry. Taking into account the time elapsed between the two data collection periods, a conversion factor of the efficiencies between the two geometries could be determined and used henceforward in the analysis. Furthermore, a natural Er target was irradiated with 7.5 MeV proton beam and via the $^{166}$Er(p,n)$^{166}$Tm reaction, a strong $^{166}$Tm source was produced, its activity was measured at both geometries, in order to verify the efficiency conversion factor derived for the 80.6 keV gamma line.

The measured $\alpha$-induced cross section values are listed in Table \ref{tab:res}. 
The effective center-of-mass energy in the second column takes into account the energy loss of the beam in the target. The quoted uncertainty in the E$_{c.m.}$ values corresponds to the energy stability of the $\alpha$-beam and to the uncertainty of the energy loss in the target, which was calculated using the SRIM code \cite{SRIM}. The uncertainty of the cross sections is the quadratic sum of the following partial errors: efficiency of the LEPS detector ($\leq$ 7\%), number of target atoms (5\%), current measurement (3\%), uncertainty of decay parameters ($\leq$ 7.8\%) and counting statistics (0.1 - 27.1\%).

\section{Discussion}
\label{dis}

The experimental results were compared to H-F calculations performed with the code SMARAGD \cite{smaragd}. In the H-F picture, a large number of resonances at the compound nucleus formation energy is described by using averaged widths containing all energetically possible particle- or $\gamma$-emission and -absorption processes. In the investigated energy range, three types of averaged widths have to be considered, in principle: the $\gamma$-, neutron-, and $\alpha$ widths. The sensitivities of the reactions are shown in Figs.\ \ref{fig:sensi_ag}, following the definition of sensitivity given in \cite{sensipaper}. The calculated cross sections are insensitive to the proton width.

Crucial for the theoretical interpretation of the data is the fact that ($\alpha$,n) data above 15 MeV and ($\alpha$,$\gamma$) data below the ($\alpha$,n) threshold have been consistently taken. At these energies, the cross sections of these reactions are only determined by the averaged $\alpha$ widths and thus their prediction in two energy regions can be unambiguously tested. On the other hand, above the neutron emission threshold the ($\alpha$,$\gamma$) cross section additionally depends on the $\gamma$ and neutron widths. The ($\alpha$,n) cross section below about 15 MeV is also sensitive to these widths. The neutron- and $\gamma$ widths cannot be determined independently, thus only their predicted ratio can be tested after accounting for the necessary $\alpha$ width modification.

\begin{figure}
\center
\resizebox{0.5\columnwidth}{!}{\rotatebox{0}{\includegraphics[clip=]{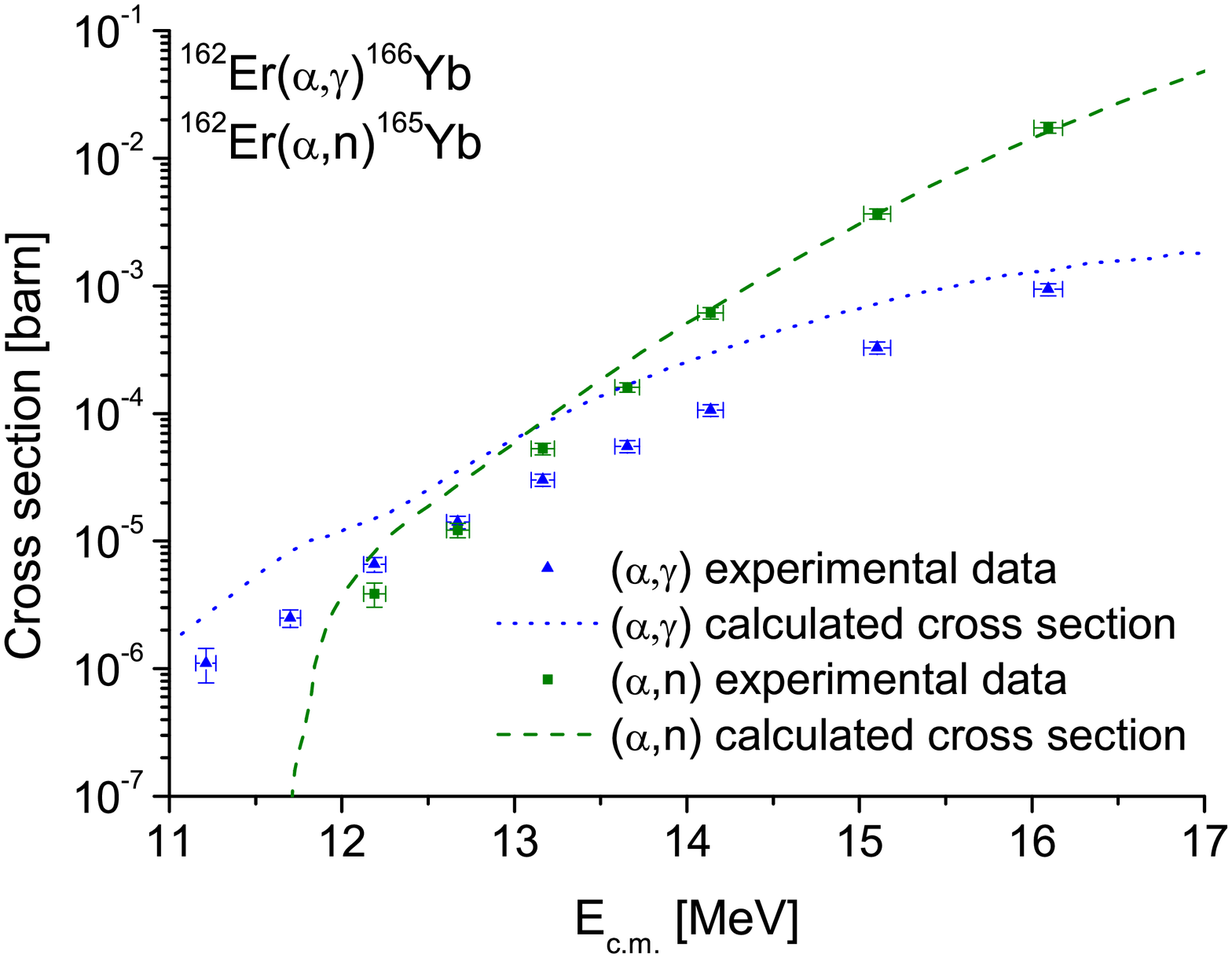}}}
\caption{\label{fig:csplot} Experimental $^{162}$Er($\alpha$,$\gamma$)$^{166}$Yb and $^{162}$Er($\alpha$,n)$^{165}$Yb reaction cross sections compared with Hauser-Feshbach calculations.}
\end{figure}

Figure \ref{fig:csplot} compares the measured cross sections of both reactions to the H-F calculations. In many cases, two types of reactions are not sufficient to constrain the averaged widths required to calculate the cross sections. Considering the sensitivities of the reactions investigated here, however, it becomes obvious that it is impossible to reproduce the data without modification of the $\alpha$ width. Moreover, the energy dependence has to be changed with respect to the one obtained with the potential by McFadden and Satchler \cite{mcf}, which was used in the default calculations shown in Fig.\ \ref{fig:csplot}, for the following reasons. From the $^{162}$Er($\alpha$,$\gamma$)$^{166}$Yb cross sections below the ($\alpha$,n) threshold it appears as if the $\alpha$ widths are predicted too large by a factor of 2.5. Scaling down the $\alpha$ width would also nicely reproduce the ($\alpha$,$\gamma$) data below 14 MeV but it would destroy the agreement of the prediction with the $^{162}$Er($\alpha$,n)$^{165}$Yb data above 15 MeV, which is only sensitive to the $\alpha$ width. Therefore the $\alpha$ width seems to be well predicted above 15 MeV but overestimated below the ($\alpha$,n) threshold. This is consistent with previously seen low-energy discrepancies between data and predictions but has never been shown unambiguously within one measurement.

\begin{figure}
\center
\resizebox{0.5\columnwidth}{!}{\rotatebox{0}{\includegraphics[clip=]{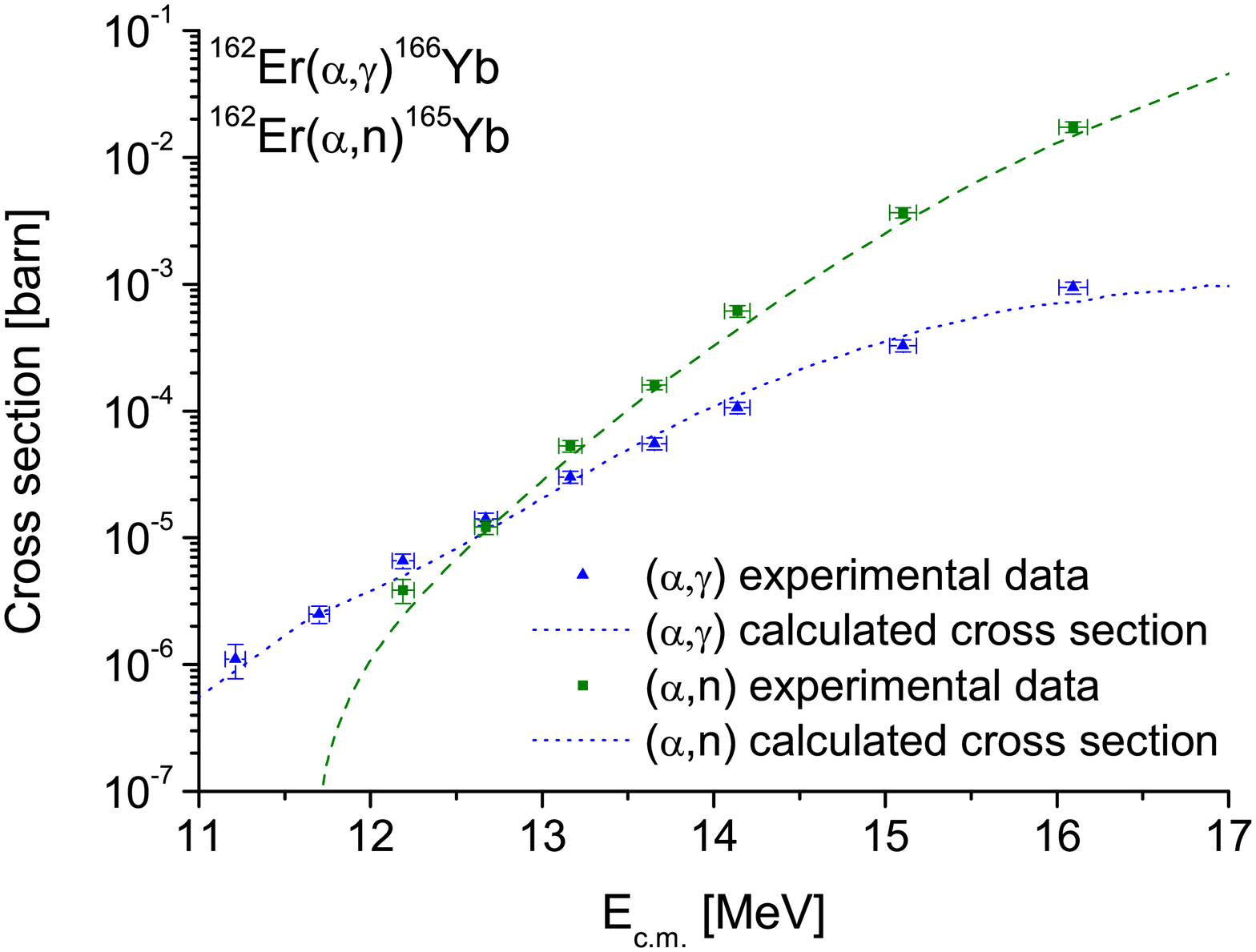}}}
\caption{\label{fig:bridgeplot} Experimental $^{162}$Er($\alpha$,$\gamma$)$^{166}$Yb and $^{162}$Er($\alpha$,n)$^{165}$Yb reaction cross sections compared with Hauser-Feshbach calculations using modified widths (see text).}
\end{figure}

Also the use of recent global, energy-dependent $\alpha$+nucleus optical potentials \cite{avri,deme} proved unsatisfactory. We were able to obtain good agreement with the data by employing the McFadden and Satchler potential \cite{mcf} but with an energy-dependent depth of the imaginary part as previously suggested \cite{somo98,sau11,deme,moh97,moh}. Similarly to \cite{sau11}, this depth $W$ is given by

\begin{equation}
W(C,E_\mathrm{c.m.}^\alpha)=\frac{25}{1+e^{\left(0.9E_\mathrm{C}-E_\mathrm{c.m.}^\alpha \right)/a_E}} \quad \mathrm{MeV},
\end{equation}

where $C$ is the height of the Coulomb barrier as introduced in \cite{avri}. Best overall agreement with the present data was found when setting $a_E=2.5$ MeV. With this optical potential for the $\alpha$ width and increasing the ratio of the neutron to $\gamma$ width by 40\%, the experimental data are reproduced well as shown in Fig.\ \ref{fig:bridgeplot}. The value of $a_E$ is close to the values used in \cite{sau11,somo98}, where $a_E=2$ MeV was used.

The present data clearly show the need for improvements in the calculation of $\alpha$-induced reactions at subCoulomb energies. We want to emphasize that the physical cause of the required modification cannot be inferred from the measurement of reaction cross sections alone, without elastic scattering data, and thus it is not possible to decide whether the optical potential has to be modified or an additional reaction channel has to be considered. In the optical model formulation of scattering theory, the (complex valued) optical potential describes elastic scattering on a scattering center \cite{blatt,satch,glen,gad,descrau,rau11}. It also determines the total reaction cross section comprised of all processes beyond elastic scattering. The standard approach in reaction calculations, also followed here, is to use an optical potential for the H-F calculations and thereby assuming that the non-elastic part of the total cross section, as given by the optical potential, is due to the formation of a compound nucleus. The H-F model is then used to predict how the total reaction cross section is distributed among the exit channels, such as the ones emitting $\gamma$-rays or neutrons. If a discrepancy between a measured channel and a prediction occurs, the cause can lie in either an incorrect prediction of the relative strengths of the exit channels or an inappropriate optical potential leading to an incorrect total reaction cross section. The latter could be tested with elastic scattering data but this is unfeasible at low energies for heavier nuclei. Nevertheless, for the case studied here below the ($\alpha$,n) threshold it seems that only the ($\alpha$,$\gamma$) channel is open and thus a modification of the optical potential is required. This view neglects the possibility of reactions which do not proceed via a compound nucleus and are not included in the H-F model. If there are such reactions, they would contribute to the total reaction cross section given by the optical potential but it would be inappropriate to use the same potential in the H-F calculation. In this context our optical potential above is an effective, modified potential to be used in the H-F calculation but not the standard optical potential as, e.g., derived from elastic scattering. This modified potential accounts for the fact that part of the reaction flux is not going into formation of a compound nucleus. In this case, were it possible to perform elastic scattering measurements at such low energies, they would not indicate a need for a modification of the optical potential. This was already pointed out in \cite{raucoulex} and the same reference suggested low-energy Coulomb excitation as such a direct inelastic reaction channel not included in the H-F formalism.

\section{Summary}
\label{sum}

The cross sections of the $^{162}$Er($\alpha$,$\gamma$)$^{166}$Yb and $^{162}$Er($\alpha$,n)$^{165}$Yb reactions have been measured for the first time at low energy. It turned out to be crucial for the theoretical interpretation that ($\alpha$,$\gamma$) data were taken below and above the ($\alpha$,n) threshold. Thus, it was possible to show consistently that either the $\alpha$+nucleus optical potential requires an energy-dependent modification or an additional reaction channel is contributing at low energies. Our conclusions support previous studies and are an important further step to improve predictions of astrophysical reaction rates for the $\gamma$-process.

\section{Acknowledgements}

This work was supported by OTKA (K101328, PD104664, K108459). G. G. Kiss acknowledges support from the J\'anos Bolyai Research
Scholarship of the Hungarian Academy of Sciences. T. Rauscher is supported by the Swiss NSF, the European Research Council, and the THEXO collaboration within the 7$^{th}$ Framework Programme of the EU. Zs. T\"or\"ok acknowledges support from the Hungarian National Excellence Program (T\'AMOP-4.2.4. A/2-11-1-2012-0001 project, financed by the European Social Fund and by the State of Hungary).   

\bibliographystyle{elsarticle-num}

\end{document}